# Adatom diffusion on vicinal surfaces with permeable steps

Bogdan Ranguelov and Ivan Markov
*Institute of Physical Chemistry, Bulgarian Academy of Sciences,*
*1113 Sofia, Bulgaria*

imarkov@ipc.bas.bg; rangelov@ipc.bas.bg

**Abstract**
We study the behavior of single atoms on an infinite vicinal surface assuming certain degree of step permeability. Assuming complete lack of re-evaporation an ruling out nucleation the atoms will inevitably join kink sites at the steps but can do many attempts before that. Increasing the probability of step permeability or the kink spacing lead to increase of the number of steps crossed before incorporation of the atoms into kink sites. The asymmetry of the attachment-detachment kinetics (Ehrlich-Schwoebel effect) suppresses the step permeability and completely eliminates it in the extreme case of infinite Ehrlich-Schwoebel barrier. The average number of permeability events per atom scales with the average kink spacing. A negligibly small drift of the adatoms in a direction perpendicular to the steps leads to a significant asymmetry of the distribution of the permeability events the atoms thus visiting more distant steps in the direction of the drift.

Vicinal crystal surfaces grow by step flow when the temperature is sufficiently high so that the mean free path of the adatoms is longer than the mean terrace width. Lowering the temperature leads to a drastic decrease of the adatom diffusivity, which gives rise to island nucleation and growth on the terraces [1, 2]. At sufficiently high temperatures the kink density along the steps is higher than the mean free path of the adatoms on the terraces and the steps represent continuous sinks for the adatoms [1]. With the decrease of the temperature the kink spacing increases and the step is no more a continuous sink. The adatoms should follow a terrace-edge-kink mechanism to join the kinks. The latter involves consecutive processes of surface diffusion on the terraces towards the steps, diffusion along the step edges and incorporation into kink sites [3, 4]. In this case atoms can leave the step before joining a kink site, or in other words, before joining the crystal lattice at least for a time. This phenomenon known as a step permeability [5, 6] was observed in a series of materials systems [7, 8, 9] and has been given a considerable theoretical attention mostly in connection with the problem of bunching of steps on vicinal crystal surfaces [5, 10, 11, 12, 13].

The growth at sufficiently high temperatures is accompanied by evaporation of atoms from the terraces. Thus an adsorption-desorption equilibrium takes place far from the steps [1] and the case is known as incomplete condensation. \cite{Ven00} At such temperatures the formation of two-dimensional nuclei on the terraces is less probable. However, at the growth temperatures usually used in experiments the re-evaporation of atoms is strongly inhibited. The reason is that the activation energies for desorption are usually of the order of the half of the heat of evaporation. We have then the case of comp\-lete condensation. At these lower temperatures the mean free path of theadatoms can become smaller than the terrace width and two-dimensional (2D) nucleation can take place. However, in most cases there is a temperature window in which the crystal surface grows by step flow and the 2D nucleation is still not possible (see Ref.[14] and the references therein).

Hence, the physics of both cases of incomplete and complete condensation significantly differ particularly in connection with the problem of the step permeability. In the former case the atoms can leave the step and re-evaporate. In the latter case the atoms have no other choice but to stay on the surface. The rate of propagation of the separate steps is given by the simple relation $v = F\lambda$ where $F$ is the atom arrival rate per adsorption site and $\lambda$ is the mean terrace width [15].
Assuming some degree of permeability in the particular case of complete condensation means that if an atom does not join a kink site at a particular step it will do that at another step. If the permeability is zero the atom will join a kink at the first step which it meets. Increasing the degree of permeability from zero onwards (and ruling out the 2D nucleation) the number of steps the atom will cross without joining a kink will increase. Some steps can be visited only once, some steps several times but at the end of its journey the atom will inevitably join a kink site.

Filimonov and Hervieu performed a detailed microscopic analysis of the elementary processes taking place at a *single* step assuming the coefficients of step permeability and impermeability are complementary [6, 16, 17]. They found that a step is permeable if it is sufficiently smooth and the edge diffusivity is low. The natural result has been a coefficient $q = \delta_0 / 2\lambda_e$ which gives the ratio of the average equilibrium kink spacing $\delta_0$ and the mean free path of the atoms adsorbed at the edge, $\lambda_e = \sqrt{D_e \tau_e}$, where $D_e$ and $\tau_e$ are the edge atoms diffusion coefficient and the mean residence time of the atoms at the step edge, respectively. When $q \gg 1$ one can expect highly permeable steps and vice versa. Decrease of the temperature does not necessarily mean increase of the permeability due to smoothing of the step. The effect of the atoms edge diffusivity should be accounted for as well.

In the present paper we are going a step further. We study the behavior of atoms on a vicinal surface consisting of infinite number of parallel steps and terraces assuming complete lack of re-evaporation and 2D nucleation. The steps are presumed to be smooth and the kinks are of one unit $a$ length. The kink spacing relative to interatomic spacing $a$ is distributed randomly by 20\% around an equilibrium average value which is produced by a thermally activated process [1]. The attachment and detachment of atoms to and from step ledges is assumed initially symmetric as a first approach to the problem. The Ehrlich-Schwoebel effects [18, 19] or the attachment-detachment asymmetry is then addressed. At the end we study the effect of a drift of the atoms on the terrace in a direction perpendicular to the steps either step-up or step-down\cite [20, 5].

We perform a Monte Carlo simulation by making use of the following model. We consider a terrace with a square lattice and a single step - Figure 1. Periodic boundary conditions provide the infinite length and the infinite number of the steps.

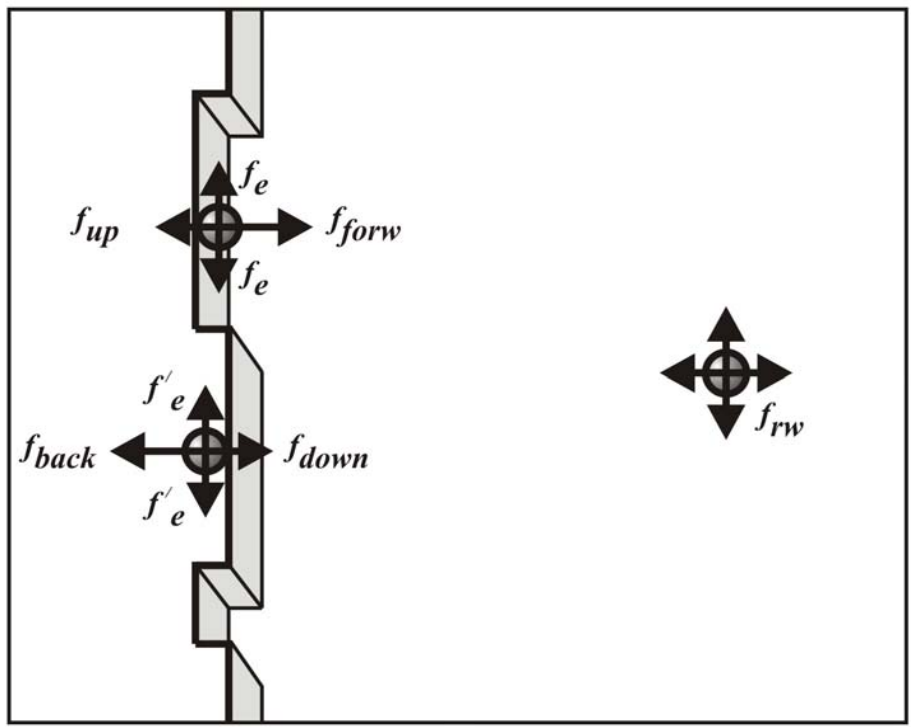

**Figure 1.** Schematic of the probabilities of different processes considered in the model

The terrace width is always much greater than the average kink spacing to ensure that atoms will meet the step at a new place which is sufficiently far from the previous one. The atoms are deposited one by one at random sites on the terrace and are allowed to walk randomly with equal probabilities $f_{rw}$ in all four possible directions. The only exception is when we allow directed drift of the atoms. Then the probability of step-up diffusion is slightly greater than the step-down diffusion (or vice versa) but the four probabilities remain complementary. The collision of atoms at the step edge and the formation of kinks by kinetic one-dimensional nucleation [21] is ruled out.

Three events are supposed to take place when an atom is adsorbed at the step ledge. It can either join a kink (impermeability event), or be reflected backward to the same terrace it came from, or to cross the step going to the neighboring terrace (permeability event). The reflection of the atom on the same terrace it has come from is not accounted for as a permeability event. We count as permeability events only the step crossings either from the upper to the lower terrace or from the lower to the upper one. Sticking of an atom to a kink site is assumed to be irreversible.

When an atom is adsorbed at the step ledge from the lower terrace four events are conceivable. The atom can either diffuse to the left and to the right along the ledge with equal probabilities $f_e$ or, it can be reflected back to the lower terrace or, it can jump on the upper terrace (step crossing) with the probabilities $f_{forw}$ and $f_{up}$, respectively. The probability $f_e$ is thus a measure of the atom diffusivity along the step ledge. When $f_e = 0$, $f_{forw} = f_{up} = 0.5$ and the step is highly permeable. In the opposite case $f_e = 0.5$ ($2f_e = 1.0$) the step is completely impermeable. If the atom moves to a neighboring site at the ledge it has again all four complementary possibilities mentioned above. The procedure is repeated until the atom either reaches a kink site (impermeability event) or leaves the step before joining a kink.

If an atom reaches the last adsorption site before the step from the upper terrace (the step rim) it can either jump down to be adsorbed at the step ledge with a probability $f_{down}$, or can go back to the same terrace with the probability $f_{back}$, or diffuse left and right along the step with the equal probabilities $f'_e$. All four probabilities are again complementary. Once the atom is adsorbed at the step ledge it has again the four probabilities mentioned above. Diffusing along the step rim the atom can also join a kink site from above [22, 23].

Figure 2 shows the distributions of the permeability events at each particular step at a constant average kink spacing and at different values of the degree of ledge diffusivity $f_e$. Symmetric attachment-detachment kinetics (absence of Ehrlich-Schwoebel effect, $f_{forw} = f_{up}$ and $f_{back} = f_{down}$) is assumed.

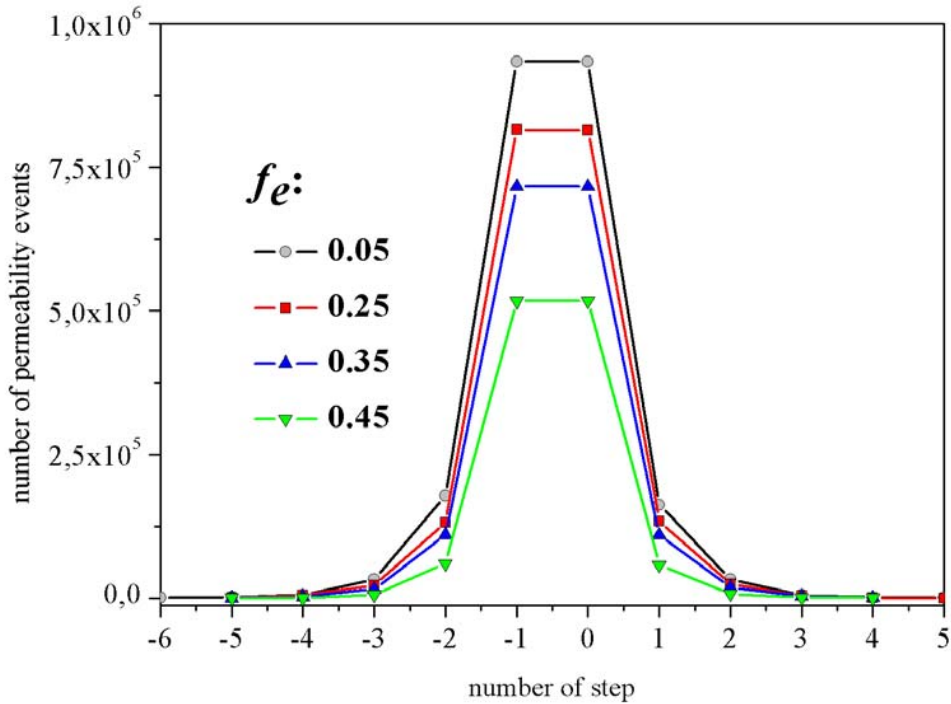

**Figure 2.** Distribution of the permeability events at each step before sticking to kink sites at different degrees of diffusivity ($f_e$) along the step ledge. The first descending step the atom meets to the right is numbered as zeroth step. The steps to the left and to the right of it are given negative and positive numbers. We count the step crossings which accumulate at each step. Mean distance between kinks is $\delta_0 = 150a$.

As seen the two neighboring steps which border the zero-th terrace on which we deposit the atoms are crossed with equal probabilities. With increasing the degree of permeability (decreasing $f_e$) more distant steps are crossed but that the probability to cross the first two steps they meet also increases. This means that the atoms cross the first two bordering steps most of the times they visit more distant steps. The effect of the average kink spacing is shown in Figure 3. The number of the crossed steps increases with increasing the step spacing at a constant degree of permeability. Again, increasing the step permeability by increasing $\delta_0$ the atoms cross more distant steps but passing through the the first two steps bordering the zeroth terrace.

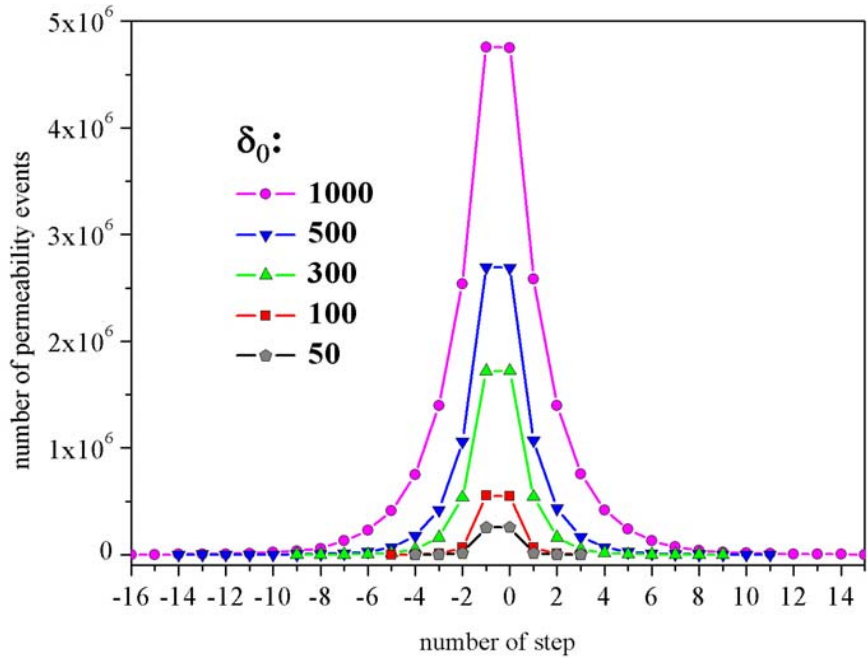

**Figure 3.** Effect of the average kink spacing $\delta_0$ on the distribution of the permeability events. Diffusivity along the step edge is $f_e = 0.2$.

Assuming asymmetry of the attachment-detachment processes at steps (Ehrlich-Schwoebel effect) decreases drastically the step permeability (see Figure 4) but the curves remain symmetric. The number of the crossed steps, including the two steps which enclose the terrace, rapidly decreases.

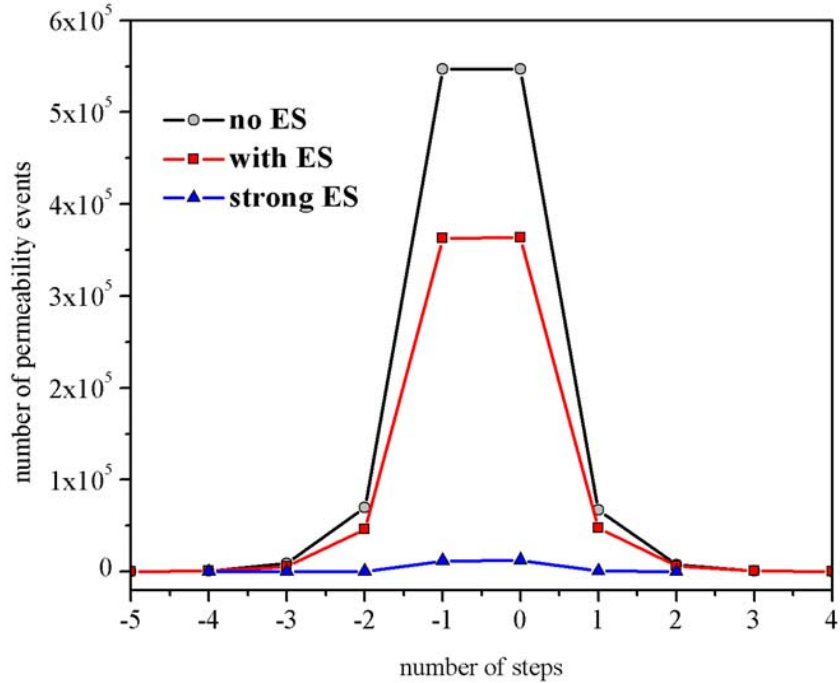

**Figure 4.** Effect of the Ehrlich-Schwoebel barrier on the step permeability. Average kink spacing is $\delta_0 = 100a$ and diffusivity along the step edge is $f_e = 0.2$.

An infinite Ehrlich-Schwoebel barrier does not allow step crossings and the steps will become completely impermeable irrespective of the mean kink spacing and the probabilities of step crossing or, what is the same the step diffusivity. The result will be the same in the case of the inverse Ehrlich-Schwoebel effect.

It is instructive to follow the behavior of the average number of permeability events per atom, $\overline{N}$, as a function of the average kink spacing, $\delta_0$, at constant probabilities $f_e$ and $f'_e$ for diffusion along the step ledge and on the step rim, respectively. We sum up all the permeability events for each distribution curve shown in Figure 3 and divide the sum by the number of atoms that have been released to perform diffusion on the vicinal surface. The result is shown in Figure 5 in double logarithmic co-ordinates. As seen perfect straight lines are obtained which is an indication of a scaling of $\overline{N}$ with $\delta_0$. At probabilities $f_e$ going to zero (complete permeability) the slopes of the straight lines become very close to each other. The inset in Figure 5 shows the slopes of the straight lines vs. the probability $f_e$. The points are fitted by a simple exponential curve.

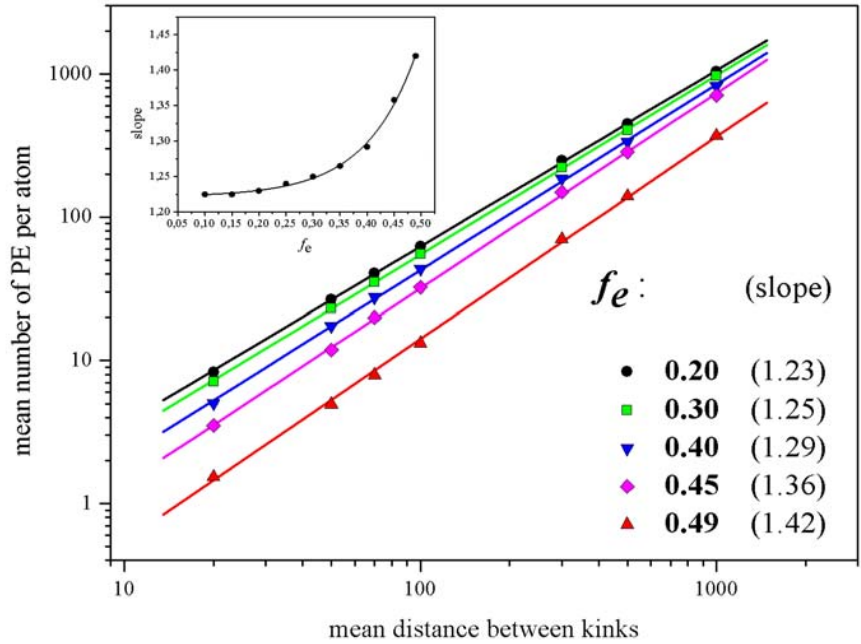

**Figure 5.** A double logarithmic plot of the average number of steps per atom permeated as a function of the average kink spacing at different probabilities, $f_e$, for diffusion along the step ledge. The inset shows the dependence of the slopes, $\chi$, on the value of $f_e$.

Figure 6 demonstrates the effect of the directed drift of the adatoms on the distribution of the permeability events. A new step has been generated after each step crossing during the calculations. The curves naturally become asymmetric but most astonishing is the fact that a negligible bias of the order of 0.001 (the probability $f_{rw}$ in a direction step-down is 0.251 whereas in a direction step-up it is 0.249) causes a dramatic shift of the curves. The curves display long tails in the direction of the drift. Increasing the bias leads to decrease of the number of crossing of the first step the atoms meet as the probability of going back is lower.

Figure 2 shows that the total number of steps crossed is in general small but the number of the permeability events is very large. The atoms prefer to go back and forth and to cross one and the same steps many times before joining a kink site. Even in the case of nearly complete permeability the atoms do not cross more than 15 steps on both sides of the zero-th terrace. The same conclusion could be made from Figure 3.

At the greatest mean kink distance of a thousand atomic spacings the atoms do not visit more than thirty steps. Thus increasing the step permeability either by decreasing the step diffusivity or by increasing the mean kink spacing leads to greater number of crossed steps. The latter is in agreement with the conclusions of Filimonov and Hervieu [6, 16, 17].

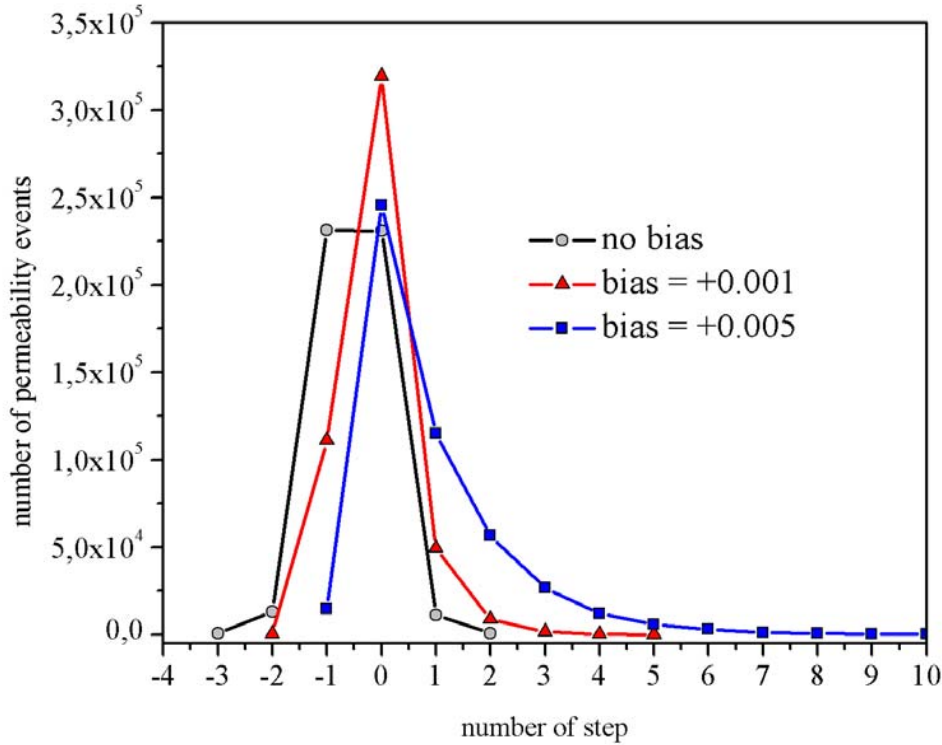

**Figure 6.** Effect of the directed drift of the adatoms on the distribution of the permeability events.

The influence of the Ehrlich-Schwoebel effect can be easily understood. The diffusion of the atoms is confined more and more in the single terrace on which they are deposited. The atoms are reflected from the ascending step and repulsed by the descending step. The atoms walk randomly on one and the same terrace and are repeatedly reflected by one and the same step until they join a kink site at the ascending step bordering the terrace. In the case of inverse Ehrlich-Schwoebel barrier the atoms will join the descending step. An infinite Ehrlich-Schwoebel barrier will tottaly eliminate the step permeability.

The curves $\bar{N}$ vs. $\delta_0$ can be described by the relation $\bar{N} = A\delta_0^\chi$ where the slope of the log-log plot, $\chi$, is an exponential function of the diffusivity of the atoms, $f_e$, along the step ledge (see the inset in Figure 5). At small values of $f_e$ the slopes $\chi$ saturate as the steps become so permeable that the atoms do not feel the difference in mean kink spacing. The case is opposite when the step diffusivity becomes very large. The steps become highly impermeable which can be compensated by larger and larger mean kink spacing.

The bunching of steps on crystal surfaces was observed for the first time in the case of Si [24]. The phenomenon was explained by a drift of the atoms in one direction owing to a force which originates from the direct current applied to heat the crystal [20]. Comparison of the experimental data with the theory showed that the force exerted on the adatoms is very small [25]. This conclusion is well compatible with the negligibly small values of the bias leading to the curves shown in Figure 6.

In summary, we studied the behavior of adatoms on a vicinal surface with permeable steps. We assumed a complete lack of re-evaporation and ruled out the two-dimensional nucleation. We have found that the atoms visit comparatively small number of steps going back and forth on the vicinal and cross multiply one and the same steps before joining a kink site. The Ehrlich-Schwoebel effect strongly diminishes the step permeability confining the migration of the adatoms to a single terrace. Very large Ehrlich-Schwoebel barriers should completely eliminate the step permeability. The mean number of permeability events per atom shows a scaling law with the mean distance between the kinks. A negligible drift of the atoms in a direction perpendicular to the steps makes the distribution curves strongly asymmetric. The curves display long tails in the direction of the drift. In any case the the step permeability should not change the rate of step propagation as the atoms inevitably join kink sites although not in the nearest step.

BR acknowledges the support of Grant No F1413/2004 from the Bulgarian National Science Fund.